# A DIY DPDK Software Router for Networking Education

Sebastian Gallenmüller
*Technical University of Munich*
Garching near Munich, Germany
gallenmu@net.in.tum.de

Stefan Lachnit
*Technical University of Munich*
Garching near Munich, Germany
lachnit@net.in.tum.de

Manuel Simon
*Technical University of Munich*
Garching near Munich, Germany
simonm@net.in.tum.de

***Abstract*—High-performance packet processing frameworks, such as the Data Plane Development Kit (DPDK), are widely used in modern network systems, yet they are rarely integrated into networking curricula. As a result, students often lack exposure to the performance considerations and low-level design choices that shape real-world packet processing systems.**

**We describe a semester-long course project in which students implement a do-it-yourself (DIY) software router using DPDK in a master-level computer networking course. The project guides students through the design and implementation of a multithreaded router, the development of a routing table data structure, and the evaluation of forwarding performance using standard benchmarking tools. We present the project structure and the virtualized infrastructure used to support classes of realistic size.**

**Based on several iterations of the course and survey feedback from participating students, we discuss the observed workload, learning outcomes, and challenges of teaching high-performance packet processing. We release the project material and scripts to recreate the infrastructure, enabling reuse and further refinement by the community.**

***Index Terms*—DPDK, Software router, Network education**

## I. Introduction

Software-based approaches are key drivers for innovation in the field of computer networks. Concepts such as OpenFlow, P4, or Network Function Virtualization allow programmability of network devices for actors outside the realm of the traditional network equipment manufacturers. This newfound flexibility and accessibility enables the rapid creation of prototypes for experimental network research and education. To provide hands-on experience with the application of Software-Defined Networking (SDN) concepts in an educational environment, several frameworks have been developed, including the network emulator Mininet [1] or the Mininet-based P4 software switches bmv2 [2] and P4Pi [3].

A prerequisite for developing software-driven approaches is the availability of high-performance platforms and frameworks. Although the previously mentioned tools provide a functional sandbox environment, their reliance on emulation [1], [2] or Raspberry Pi hardware [3] was not intended to deliver the performance required for real-world scenarios. Important factors, such as multicore scalability or other low-level optimizations, are not reflected in these frameworks, creating a gap between what is taught in networking classes and what is relevant in real-world contexts.

Our key educational contribution demonstrates how production-grade packet processing can be integrated into a scalable university course without sacrificing realism or manageability. Therefore, we designed a project combining software-based packet processing with performance-relevant aspects. The semester-long course project is offered alongside the master-level lecture *Advanced Computer Networking (ACN)* at the Technical University of Munich, complementing the theoretical concepts of the lecture with practical implementation and evaluation. The target audience for ACN comprises computer science students with prior knowledge of computer networks, who have chosen computer networks as a focus area. As part of this focus area, ACN covers selected topics in depth, in particular, network measurements, performance analysis, and programmable networking. The software router project has been designed as a challenge for students in this focus area, introducing them to the topics, tools, and techniques actively used in networking research. The clearly targeted audience allowed us to make design decisions for the router project without compromises or abstractions used for a network fundamentals course for inexperienced students. In the course project, students implement a software router using the Data Plane Development Kit (DPDK), an established framework for high-performance packet processing. Routing is a core networking function, but sustaining throughput at millions of packets per second requires addressing challenges in scalable software design, low-level Layer-2 and Layer-3 protocols, and performance-critical routing data structures. Together, these challenges make routing a non-trivial and educational problem for students.

In this work, we aim to achieve the following educational goals: (1) enable students to design and implement a complete software router in a one-semester course project, applying important concepts taught in a master-level course on computer networks, such as SDN or high-performance packet processing; (2) expose students to production-grade packet processing by working with the DPDK framework, allowing them to apply performance-critical design decisions, involving multithreaded architectures, memory management, or multiqueue NICs, which are typically abstracted away in other educational exercises, limiting students' exposure to real-world performance trade-offs; (3) create an infrastructure for the students to develop and test their software routers in a real-world environment that scales for course sizes with 30 students or more; (4) foster students' practical system skills by enabling them to implement low-level protocol processing and routing data structures, and evaluate forwarding performance using standard benchmarking tools.

## II. Related Work

We review related work on the definition of educational goals and teaching environments for networking education.

Bloom et al. [4] introduced a framework for categorizing educational goals, which was adapted for Computer Science (cf. joint ACM/IEEE Computing Curricula 2020 task force [5]). In this work we use the modified six-level taxonomy: *Remembering*, *Understanding*, *Applying*, *Analyzing*, *Evaluating*, and *Creating*.

Mininet [1] is a network emulator used for teaching. It offers a Python API to create complex network experiments on a single Linux machine. Mininet has been used for reproducing experiments of others [6] or teaching network routing [7] and SDN [8]. The mentioned examples provide valuable insights into how problems can be solved from a functional perspective. However, emulation cannot provide realistic performance, leading to significant differences between the teaching and real-world environments. P4Pi [3] is a software switch designed to teach packet processing using P4. P4 is a language for expressing packet processing problems, offering specialized data structures, e.g., for routing. P4Pi runs on top of DPDK; however, many aspects of the backend are abstracted through P4. We see significant value in teaching and understanding these low-level concepts, a gap that we want to close with this work. A P4 router provides insights into packet headers, but concepts like longest prefix matching (LPM) and its implementation are completely hidden from the user. Other educational studies [7], [8] approach routing from a higher level. There, students must provide the correct configuration to forward packets correctly, but do not gain deeper insight into the actual forwarding process.

We aim to give students full access to a realistic setup for developing and implementing the software router, based on a virtualized rather than an emulated environment, relying on typical Linux command-line tools. We avoid hiding or abstracting packet processing tasks. Students gain access to the full capabilities of the DPDK framework, implementing typical packet-processing steps, including protocol parsing, conformance checks, protocol implementation, and the creation of efficient data structures.

## III. Infrastructure

In this section, we outline the project's requirements and describe the hardware and software infrastructure that will support it.

### A. Requirements

In the following, we describe the infrastructure requirements to meet all desired learning objectives.

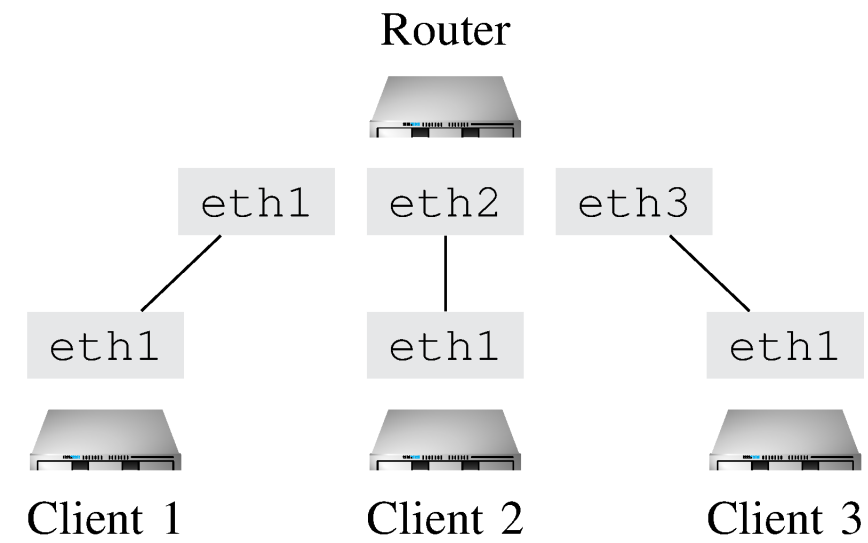

Fig. 1: Minimal router topology

*Topology:* Our topology consists of four nodes: the router, executing the software router, and three client nodes, to test the router (cf. Figure 1). Three clients are necessary to test the forwarding decisions, e.g., if traffic from Client 1 is correctly distributed between Clients 2 and 3. The router node has three interfaces (`eth1`–`eth3`), one interface per client. These connections form the *experiment network* for developing and testing the software router. Router and client nodes have a separate management interface (hidden in Figure 1) attached to a *management network*. The isolated management network guarantees students continuous node access, independent of errors or misconfigurations in the experiment network.

*Scalability:* To keep the entry barriers for students low, we provide a development and test setup for each participant. However, providing four bare-metal machines for every participant is infeasible, e.g., a master-level course with 30 students would require 120 nodes in total. Node virtualization is a solution that supports the anticipated scale.

*DPDK compatibility:* Our goal is to provide a realistic setup and experience for using DPDK. DPDK ships with its own drivers and requires compatible NICs to fully utilize the performance and capabilities of the available network hardware. We choose emulated `virtio` NICs[1], which are directly supported by DPDK's drivers, offering an identical setup to non-virtualized NICs. The DPDK framework requires a more complex setup than other widely-used frameworks, like Mininet, or the Linux-based solutions, such as XDP or eBPF. However, DPDK provides only limited abstraction and low-level access, two relevant requirements for the planned learning outcomes. The more complex setup also provides additional insights for participants into another relevant topic, such as virtualization.

### B. Setup

Below, we describe the setup used for our lecture fulfilling our requirements.

A single instance of the described topology requires 6 CPU cores (1 core per router interface, 1 core per client) and 3.5 GB of RAM (2 GB for the router, 512 MB per client). The total resource consumption for an expected course size of 30 students is 180 CPU cores and 105 GB of RAM. However, average resource consumption will be lower, because only a limited number of students will use their VMs simultaneously. For our lecture, we use a workstation running a 64-core CPU (AMD EPYC 7763) with 1 TB of RAM.

[1] https://doc.dpdk.org/guides-24.11/nics/virtio.html

We create a git repository for each student, where they can submit their solutions to the assignments. In addition, we provide detailed instructions [9] and a Vagrant script [10] to deploy the 4-node topology in an OS-independent way. The Vagrant scripts are well-tested on our workstation. Therefore, we recommend using the workstations (accessible via ssh) for less-experienced users. More experienced users can use the Vagrant script to host the topology on their own machines.

## IV. Project and Assignment Design

In this section, we introduce the structure of the course project, divided into four problems (*P1–P4*) covering seven learning goals (*LG1–LG7*). Section VIII describes all released artifacts. We design the project to cover different levels of Bloom's Taxonomy. The lecture covers the first two levels, *remember* and *understand*, the practical assignments focus on the higher levels. We divide the learning goals into separate project tasks (cf. Table I). *P1* familiarizes students with the involved tools and setups, *P2* focuses on implementing basic router operations. Afterward, students implement a high-performance routing table (*P3*), and benchmark the software router (*P4*).

### A. Problem 1: Fundamental Tools and Setup

In *P1*, students learn to use basic command-line tools to prepare for the subsequent problems (LG1). We purposely selected well-known tools; therefore, the students do not work in a simplified exercise environment, but in a realistic Linux environment. As the VMs do not include a GUI, students learn to use the command line, a skill that we deem highly important. To perform the tasks of all problems, students use `ssh` to access and configure the VMs, `git` to save and submit the scripts containing their solutions, and `ping` and `tcpdump` to test the network setup.

For the next tasks, students configure the NICs of the experiment network in two different ways: (1) on the client interfaces via `ip` and (2) on the router interfaces via the DPDK tools. For DPDK, all necessary steps are listed in a readme file of the provided framework [11], containing an unmodified DPDK (v24.11 LTS). The only difference between the official and the exercise framework is that we included additional files for implementing the software router. Students submit shell scripts that contain the steps to configure client and router nodes on a fresh VM setup.

Finally, the students test their scripts using `ping` between Clients 1 and 2 while the Router runs a DPDK forwarder application. Students observe the network using `tcpdump` and submit their findings.

### B. Problem 2: Design and Implementation

In *P2*, the core functionality of the router is implemented that handles the forwarding of IPv4 packets. Students learn to apply foundational protocol concepts (LG2); they extract protocol information from an RFC and implement Layer 2 and 3 protocols from scratch using DPDK.

A major factor for the high performance of DPDK applications is the multithreaded architecture (LG3). For maximum performance, synchronization between different threads should be avoided. Multi-queue NICs offer separate receive and transmit queues for each thread, avoiding the overhead of synchronizing NIC access between threads. We introduce these architectural concepts in the lecture and suggest the following architecture. The router consists of three threads, one thread per interface. Each thread manages a single receive queue of its associated interface. Each thread further manages three transmit queues (one per interface), ensuring that it can send out packets on all available interfaces. Multithreaded architectures, avoiding synchronization, and end-to-end packet processing on a single thread are not unique to DPDK but general concepts for achieving high performance in packet processing.

As the routing table will be implemented in *P3*, we provide a dummy routing table with hard-coded forwarding decisions.

RFC 1812 [12] defines basic operations and sanity checks for the IPv4 header that routers must perform. Students must extract these checks from the RFC and implement them in their router. To keep the software router complexity manageable, we do not require the students to implement ICMP.

To avoid manually configuring MAC addresses on the clients, the router needs to respond to ARP requests directed at its own interfaces. After integrating ARP, the router can reply to ARP requests of the clients and forward ping messages between clients.

Apart from the routing table interface and the command-line arguments, the students have considerable freedom to implement the router. This resembles a realistic environment with a limited number of well-defined constraints and a typically wider space where behavior is more loosely defined (e.g., correct memory handling) and requires the students to test their implementation (LG5). Solving problems in such an environment is a highly valuable learning outcome of the course project.

### C. Problem 3: Routing Table

We use the DIR-24-8 [13] algorithm for the routing table. The lecture introduces this algorithm, equipping students with the necessary background for implementation. In this problem, students learn to apply the concepts introduced in the lecture to create a low-level data structure (LG4).

The provided framework includes a header file defining the routing table interface and rudimentary tests. This allows students to test their implementation. However, the number of given test cases is limited intentionally. We want to avoid students adapting to a purely test-dominated implementation process, where the implementation is minimally changed, compiled, and tested. A crucial part—reflecting on the changes made—can be missed in such a process. We encourage students to extend the tests, adding their own test cases.

TABLE I: Mapping of Learning Goals to Problems 1 to 4 (*P1–P4*) and Bloom Levels

| | Learning goal | Bloom Levels | *P1* | *P2* | *P3* | *P4* |
|---|---|---|---|---|---|---|
| LG1 | **Use** packet processing and networking tools | *Apply* | ✓ | | | ✓ |
| LG2 | **Analyze** and **implement** fundamental protocol processing from RFC specifications | *Analyze/Create* | | ✓ | | |
| LG3 | **Apply** and **optimize** performance and design principles for network systems | *Apply* | | ✓ | ✓ | ✓ |
| LG4 | **Design** and **implement** low-level data structures for high-performance networking | *Create* | | | ✓ | |
| LG5 | **Develop** and **evaluate** suitable test cases to ensure correctness | *Evaluate/Create* | | ✓ | ✓ | |
| LG6 | **Define**, **measure**, and **analyze** benchmarking KPIs | *Apply/Evaluate* | | | | ✓ |
| LG7 | **Visualize**, **interpret**, and **evaluate** experiment results | *Evaluate* | | | | ✓ |

This way, students learn to design and implement effective test cases for a specific implementation (LG5). Low-level, hardware-optimized programming is prone to implementation errors, e.g., out-of-bound accesses. Tests that check these edge cases help detect these errors. Finding these edge cases and creating tests are highly valuable skills taught by this exercise.

DIR-24-8 requires programming bit-level operations. These low-level implementation tasks are often neglected in university courses focusing on high-level concepts or software engineering (LG3, LG4). Implementation exercises in low-level data structures help broaden students' skill sets.

### *D. Problem 4: Benchmarking Software Routers*

In this problem, students execute and evaluate a benchmark to measure the throughput of the software router created in *P2* and *P3*. The benchmark is inspired by RFC 2544 [14], investigating the throughput performance for network interconnect devices. Students learn to perform and evaluate scientific measurements using standard tools (LG1, LG6).

For benchmarking, we use the packet generator `iperf3`. The packet generator is executed on Clients 1 and 2 to measure the forwarding performance of the router at specified packet rates and packet sizes. Based on the measurements, students create plots of the measured ingress and egress packet rates. For the evaluation, we provide a `Jupyter` notebook that uses the plotting library `matplotlib`, two established tools in research. As a final task, students briefly describe their observations and discuss which factor is more important, the rate or size of packets (LG3, LG7).

### *E. Pedagogical Design Considerations*

Students have approx. 4 weeks to prepare each problem and receive a maximum of 10 credits for successful participation in the course project. The number of credits awarded reflects the difficulty level and the time required to solve the respective problems. *P1* (max. 1 credit) is designed to give a low-barrier entry to the project. *P2* and *P3* (max. 4 and 3.5 credits) require students to design and implement the software router and routing table, representing the core of the project (75 % of the achievable credits). *P4* (max. 1.5 credits) is comparatively brief and does not require the implementation of low-level code. It is intentionally kept shorter than the other problems, as students typically have less time during the end-of-semester exam period. *P4* includes a short survey on the project (cf. Section V). Students receive 0.5 credits for participation in the survey to increase the number of replies.

To maximize participation in the course project, we award a grade bonus for the final exam for successful participation. The router is designed as a challenging project, i.e., very few students will correctly solve all problems. To encourage further participation, we award the maximum grade bonus for the final exam if $\geq 75\,\%$ of the maximum credits for the project have been reached.

Typically, not all students solve each problem for various reasons. As subsequent problems of the course project require solutions to previous problems, this may lead to unnecessary drop-outs. To avoid that, we provide re-entry possibilities for each problem. *P2* can be solved with a short recap of *P1*, based on the solution presented in the lecture. The routing table in *P3* can be written based on tests provided by our framework, even without a working router. Students can benchmark the Linux router in *P4* if they did not solve *P2* and/or *P3*.

For the course project, we intentionally limited the functionality of the router: it only supports the minimum subset of ARP/IPv4, and no ICMP. These simplifications reduce complexity to match the one-semester time frame without limiting the learning objectives. Essential functionality, such as the high-speed parallel architecture, header parsing, LPM, etc., remains part of the router. These concepts are applied only to a limited number of protocols, reducing implementation effort. The omitted protocols are still part of the curriculum, so students will learn the necessary theoretical background.

DPDK is written in C, making C or C++ the obvious choice for implementing the router. Both languages allow low-level control over hardware and memory management, essential features for performance-critical applications. Using C and C++ may deter students from participating in the project. Therefore, we initially thought to provide an abstraction layer on top of DPDK to use an easy-to-learn scripting language for the router. This was already done for projects such as MoonGen [15] that use the Lua scripting language for implementing packet processing applications on top of DPDK. We see two disadvantages of using a scripting language: (1) the router will deviate from systems used in real-world scenarios, and (2) some problems will be harder to solve using a scripting language. One of these problems is the routing table, which requires bit-level operations. Scripting languages with dynamic typing require additional measures, such as explicit casting or additional functions to implement these operations. This diversion from the typical way to use these scripting languages makes implementation harder than initially anticipated. Using C and C++ for the implementation of high-performance data structures is a more natural choice, without requiring the previously described workarounds for scripting languages.

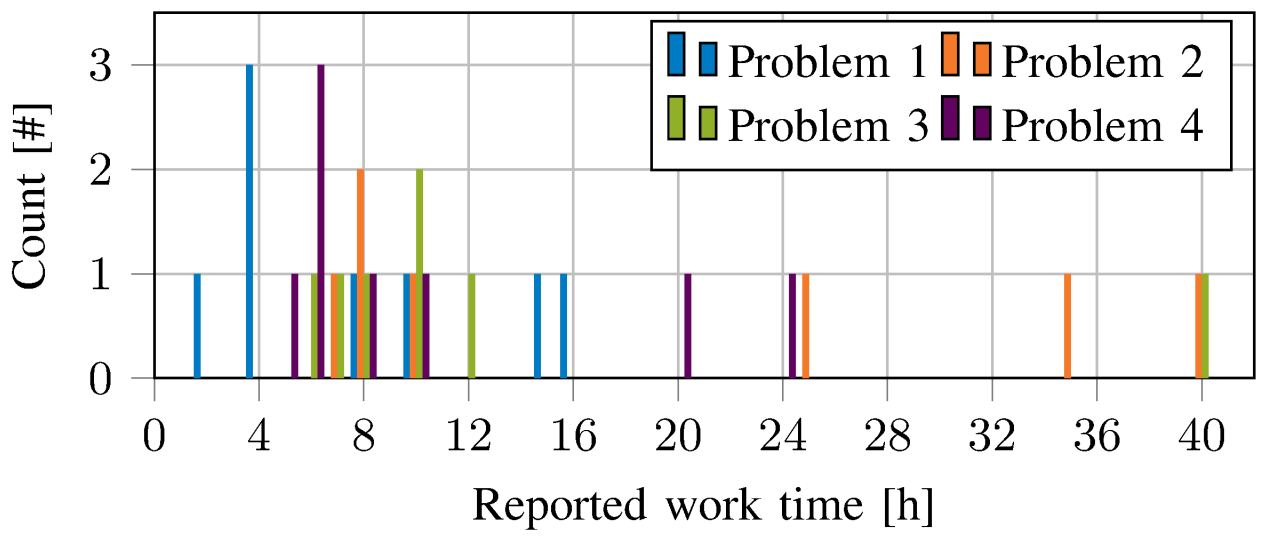

Fig. 2: Reported work times

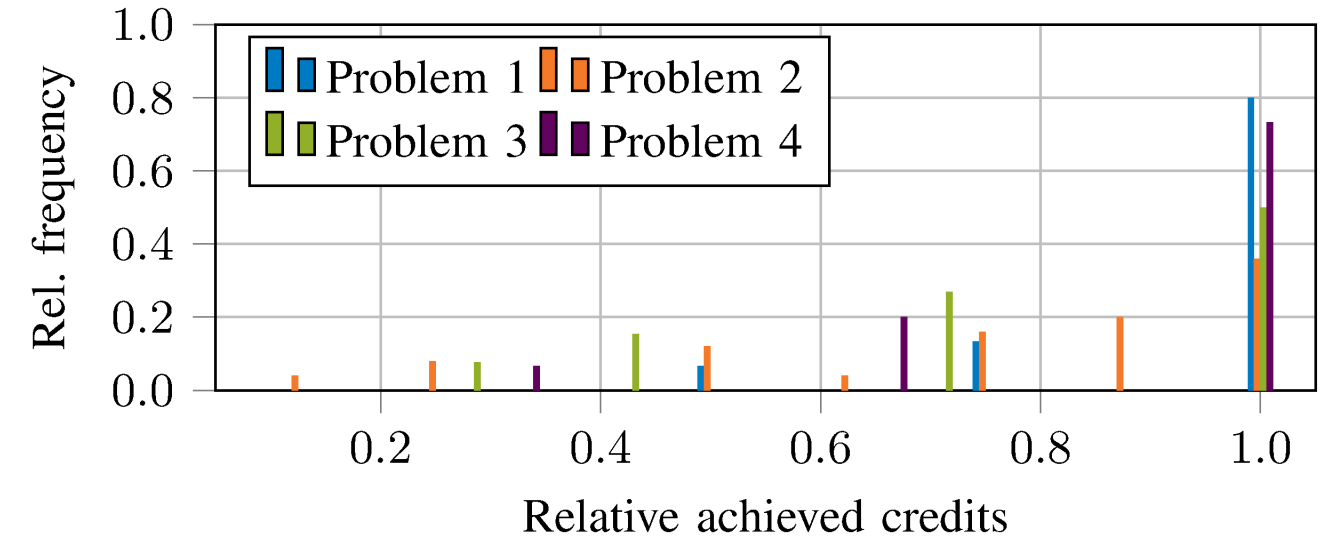

Fig. 3: Awarded credits

### *F. Grading*

After submission, students get feedback in two ways: (1) After the submission deadline of a problem, we present a solution concept in the lecture. This presentation includes typical errors made in previous years and possible solutions. (2) The submissions are individually graded by teaching assistants. Their grading includes a detailed explanation of errors that helps students solve the subsequent problems of the project. Grading typically takes a teaching assistant approximately an afternoon per problem. We rely on manual checks for *P1* and *P4* as the solution space for the scripts (*P1*) and graphs (*P4*) is limited. Therefore, wrong answers can be detected easily, making grading efficient and fast. For *P2* and *P3*, we use automated tests, checking the correct handling of ARP and IPv4 packets and the routing table.

## V. Survey Evaluation

To collect feedback on the course project, we include a short survey at the end of the project. The results were collected during the winter semesters of 2023 and 2024. 35 students participated in the router project, 12 students submitted answers for the survey. However, some students only solved a subset of the four problems, and some students did not answer all questions. Given the comparatively low sample size, the following observations should be considered indicative rather than fully conclusive.

We asked the students how long it took them to solve the problems (cf. Figure 2). Times were reported by the students themselves, involving biases and approximations. However, the numbers show rough trends in how much effort needs to be invested to solve each problem. *P1*, with a median reported work time of 6 h, could be solved with less effort than the other problems. This meets our expectations to give students a lower barrier to entering the project. For *P2* and *P3*, students reported a median work time of 10 h, and for *P4*, a median work time of 7 h. These numbers reflect our expectations: implementing the router requires more effort than benchmarking it. These work times also indicate that the current distribution of credits between the problems is justified. Figure 3 shows the awarded credits per problem. *P1* and *P4* seem to be easier, where >76 % of students received the maximum amount of credits. For *P2* and *P3*, we observe many partially correct submissions, suggesting that these problems are harder to solve.

We also asked the students to describe what they liked and disliked most about the course project. We categorized the answers and the two most liked themes were the *hands-on experience* (mentioned 5 ×) and the *concept and topic* (mentioned 4 ×) of the course project. The two most disliked themes concern the *infrastructure* (mentioned 5 ×) and the *documentation* (mentioned 5 ×). These answers demonstrate a hard-to-solve trade-off. On the one hand, students liked the hands-on experience of designing a software router from scratch with the different tools and systems involved. On the other hand, they disliked debugging the complex setup and would like to have more examples for scripts or DPDK apps, including documentation, or more information in the lecture. We see both sides of this coin inherently linked: interesting problems tend to be complex and require more effort to solve. Developing these problem-solving skills is a core competency that we want to teach the students. All of the tools and frameworks used, including DPDK, are well-documented. Including exercise-specific documentation or introducing details in the lecture would help students solve the specific problems faster and more efficiently. However, finding proper documentation is a highly valuable skill that we can teach our students. Therefore, we are cautious when creating documentation material specifically for the course project. Based on their feedback, we assume that students are not used to Linux man pages or DPDK documentation, therefore, we plan to improve the instructions. We intend to include recommendations on where and how to find the documentation material. The documentation of lab equipment is currently being improved. We aim to provide more comprehensive documentation and examples of how to use it for future iterations of the course project.

## VI. Lessons Learned

Based on the achieved credits (cf. Figure 3), we see that students successfully participated in the project. Participation numbers remained at approx. the same level for *P1–P3* (30 to 25 submissions), only for *P4*, participation dropped (15 submissions). We attribute this drop to the start of the exam period. These findings indicate that we accomplished the initially defined educational goals. Students successfully implemented a software router, using production-grade DPDK on a real-world infrastructure, while developing practical system skills.

We were surprised that the survey did not contain any complaints on the requirement to use C or C++ as a programming language for the project, given the rather limited exposure to them during the bachelor-level courses.

In the winter semesters of 2023 and 2024, 11 out of 35 students who participated in the course project used the possibility to re-enter the course project. Therefore, we highly recommend a course design that allows this re-entry.

Before 2023, students could not restart or restore their VMs. Students occasionally managed to lock themselves out of their VMs, e.g., by removing the default route. To regain access, they had to contact a teaching assistant to fix the issue. This led to a significant delay for them to continue. Therefore, we introduced the setup scripts that the students control themselves, i.e., in case of an error, they can delete their VMs and recreate them using the setup scripts. To avoid complaints and reduce the load on the teaching assistants, we highly recommend providing students with reset and restore capabilities.

## VII. Use of AI

The original project was designed before the rise of large language models (LLMs). Nowadays, LLMs can solve all described exercises with consistently high grades. There is currently no practical way to reliably detect or prevent LLM usage. To still assess students' rather than LLMs' performance, we apply additional measures.

To incentivize participation in the router project, students can earn bonus credits for the closed-book exam at the end of the semester. The bonus is only applied to the final grade if the students achieve a passing grade without the bonus. In this closed-book exam, students must demonstrate their abilities without the help of books, lecture material, and, most importantly, without access to LLMs. This ensures that only students who demonstrate their learning in the exam benefit from the grading bonus. Using such a controlled, final assessment empowers students to use LLMs responsibly. LLMs can be used, e.g., to support their coding. However, it prevents them from passing the course without understanding the central concepts taught.

## VIII. Reproducibility

We provide all necessary artifacts for others to offer the software router as part of their lecture, including a detailed description of the entire project [9], the code to set up the infrastructure [10] and the DPDK framework [11].

We plan to continue offering the course project in upcoming semesters and aim to provide a fair opportunity for future participants. Therefore, we do not provide the material used for grading.

## IX. Conclusion

Here, we presented a well-structured course project where students gain hands-on experience, designing and developing a software router during a one-semester master's course in computer networking. Compared to other teaching-focused projects that prioritize simplicity and accessibility, the described project includes performance optimization as a central learning outcome. Therefore, we chose DPDK, a high-performance packet processing framework, as our target platform. We also describe an infrastructure that can host a DPDK-compatible, virtualized development environment for an entire master course (approx. 30 students). Students will learn valuable skills solving challenging problems in a production-grade, virtualized environment using real-world frameworks and tools.

We release all artifacts to replicate the course project [9]–[11]. We hope that the community draws inspiration from our described course project and the released artifacts, extending and continuing our development.